\documentclass[twocolumn,showpacs,preprintnumbers,superscriptaddress,prl,amsmath,amssymb,nofootinbib]{revtex4-1}

\usepackage{graphicx}
\usepackage{dcolumn}
\usepackage{bm}
\usepackage{color}
\usepackage{ulem}
\usepackage{gensymb}
\usepackage{braket}
\usepackage{epstopdf}
\usepackage{amsmath}
\usepackage[percent]{overpic}

\newcommand{\MBT}{\,MnBi$_2$Te$_4$}
\newcommand{\MST}{\,MnSb$_2$Te$_4$}
\newcommand{\MBST}{\,MnBi$_{2-x}$Sb$_x$Te$_4$}

\begin{document}

\title{Vapor transport growth of \MBT~and related compounds}

\author{J.-Q. Yan}
\affiliation{Materials Science and Technology Division, Oak Ridge National Laboratory, Oak Ridge, Tennessee 37831, USA}
\email{yanj@ornl.gov}

\author{Z. L. Huang}
\affiliation{Department of Physics and Astronomy, Rutgers University, Piscataway, New Jersey 08854, USA}

\author{W. D. Wu}
\affiliation{Department of Physics and Astronomy, Rutgers University, Piscataway, New Jersey 08854, USA}

\author{A. F. May}
\affiliation{Materials Science and Technology Division, Oak Ridge National Laboratory, Oak Ridge, Tennessee 37831, USA}

\date{\today}

\begin{abstract}
Motivated by fine tuning of the magnetic and topological properties of \MBT~via defect engineering, in this work, we report the crystal growth of \MBT~and related compounds using vapor transport method and crystal characterization  by measuring elemental ratio, magnetic and transport properties, and scanning tunneling microscopy. For the growth of \MBT~single crystals, I$_2$, MnI$_2$ , MnCl$_2$, TeCl$_4$, and MoCl$_5$ are all effective transport agents; chemical transportation occurs faster in the presence of iodides than chlorides. \MBT~crystals can be obtained in the temperature range 500$^\circ$C-590$^\circ$C using I$_2$ as the transport agent. We further successfully grow \MST, \MBST, and Sb-doped MnBi$_4$Te$_7$ crystals.  A small temperature gradient $<$20$^\circ$C between the hot and cold ends of the growth cmpoule is critical for the successful crystal growth of \MBT~and related compounds. Compared to flux grown crystals, vapor transported crystals tend to be Mn stoichiometric, and Sb-bearing compositions have more Mn/Sb site mixing. The vapor transport growth provides a new materials synthesis approach to fine tuning the magnetic and topological properties of these intrinsic magnetic topological insulators.
\end{abstract}

\maketitle

\section{Introduction}

\MBT was intensively investigated in the last few years after it was first predicted and investigated as the first intrinsic antiferromagnetic topological insulator\cite{otrokov2019prediction}. \MBT~is one member of a large family of compounds with a general formula $n$MnTe.$m$Bi$_2$Te$_3$ where both integers $n$ and $m$ can vary. $n$ determines the thickness of the magnetic slabs, and $m$ determines the nonmagnetic spacing between the magnetic slabs. Varying m while keeping $n$=1 gives \MBT ($m$=1), MnBi$_4$Te$_7$ ($m$=2), MnBi$_6$Te$_{10}$ ($m$=3), and MnBi$_8$Te$_{13}$ ($m$=4) with various magnetic and topological properties which were the focus of the community in the last three years. Recently, Mn$_2$Bi$_2$Te$_5$ ($n$=2, $m$=1) was successfully synthesized\cite{cao2021growth} as the first $n>$1 member in this large family. The Sb analogues, \MST~and MnSb$_4$Te$_7$, were found to exist\cite{yan2019evolution,huan2021multiple} which facilitates the fine tuning of the physical properties via chemical substitution\cite{yan2019evolution,chen2019intrinsic,hu2021tuning,chen2021coexistence,shao2021pressure}. This family of compound provides an ideal materials playground for the study of intimate coupling between magnetic and electronic topological properties and for exotic quantum phenomena.

However, this family of compound,  like other crystalline solids, has defects in its crystal lattice that can affect both magnetic and topological properties. Density functional theory calculations suggest that the formation of Mn/Bi site mixing in \MBT~is driven by the lattice mismatch between Bi$_2$Te$_3$ and MnTe slabs\cite{du2021tuning}. The nonstoichiometry of Te and the site mixing between Bi and Te should also be considered depending on the growth conditions. These are confirmed by the experimental observation of lattice defects via single crystal x-ray and neutron diffraction, scanning tunneling microscopy(STM), scanning transmission electron microscopy(STEM) \cite{yan2019crystal, yuan2020electronic, zhu2020investigating, li2020antiferromagnetic, zeugner2019chemical, murakami2019realization, liu2021site, huang2020native, garnica2021native, shikin2021sample, sitnicka2021systemic}. We noticed in our flux-grown \MBT~crystals, while the concentration of Mn$_{Bi}$ (Mn at Bi site) is in the range of 2-4\% in crystals from different batches, the concentration of Bi$_{Mn}$ (Bi at Mn site) can vary in a wide range 4-15\% which is rather sensitive to growth parameters. In \MST, more Mn/Sb site mixing can form due to the smaller size difference between Mn and Sb. These lattice defects induce electronic inhomogeneities, a complex ferrimagnetism in each septuple layer, and can affect and even change the sign of inter-septuple-layer magnetic interactions.\cite{lai2021defect, murakami2019realization, liu2021site, huang2020native}  Recently, lattice disorder in \MBT~was proposed to account for the experimentally observed high Chern number state in \MBT~ flakes\cite{li2021coexisting} and affect the gap size of the surface states\cite{garnica2021native, shikin2021sample, sitnicka2021systemic}. Careful investigations on MnTe.$m$Bi$_2$Te$_3$ with $m>$1 found the above mentioned lattice defects in both the quintuple and septuple layers, which lead to diverse magnetic ground states\cite{yan2021delicate,hu2021tuning}.

The presence and abundance of lattice defects make possible the defect engineering of the magnetic and topological properties of this family of compound. Considering the sign change of the inter-septuple-layer magnetic interactions with increasing amount of Mn$_{Sb}$ (Mn at Sb site) in \MST\cite{murakami2019realization, liu2021site}, similar approach might lead to ferromagnetic \MBT~with either increasing amount of Mn$_{Bi}$ or partial substitution of Bi by other transition metal ions such as V or Cr. This ferromagnetism will facilitate the realization of quantum anomalous Hall effect (QAHE) if band inversion preserves in the presence of high concentration of lattice defects. On the other hand, reducing the defect concentration to be minimal is necessary for the intrinsic properties and possibly quantized transport properties at even higher temperatures. Thermodynamically, materials syntheses at lower temperatures helps reduce the amount of site mixing while growths at higher temperatures are expected to have an oppositive effect. The different growth mechanism and kinetics of different growth approaches can be employed to fine tune the lattice defects and nonstoichiometry and therefor the physical properties.

In the pioneer work by Otrokov et al\cite{otrokov2019prediction}, \MBT~single crystals were grown by slow cooling of a  melt mixture of MnTe and Bi$_2$Te$_3$ in a narrow temperature window or by Bridgman method. Later, we developed the flux growth of sizeable \MBT~crystals out of Bi$_2$Te$_3$ flux and this procedure can also grow other members in the MnTe.$m$Bi$_2$Te$_3$ family including the Sb-doped compositions\cite{yan2019crystal}. The identification of Bi$_2$Te$_3$ as an appropriate flux for the growth of sizable MnBi$_2$Te$_4$ single crystals facilitates a thorough investigation of this interesting compound by multiple probes including inelastic neutron scattering which normally requires a large mass of samples. A careful inspection of MnBi$_2$Te$_4$ single crystals grown using this protocol found about 2-4\% Mn$_{Bi}$ and 4-15\% Bi$_{Mn}$; both depend on the growth parameters  especially the temperature profile and the composition of starting materials. According to the Bi-Te binary phase diagram, varying the Bi/Te ratio in the flux can slightly lower the growth temperature and can also fine tune the charge carrier concentration\cite{du2021tuning}. However, once the flux composition is too far away from Bi$_2$Te$_3$, other competing phases precipitate. This leads to limited variation of growth temperature and forces us to consider new fluxes or other growth techniques such as chemical vapor transport. Vapor transport growths are normally performed in a sealed ampoule and chemical transportation occurs with the aid of a volatile substance named transport agent. At elevated temperatures, the transport agent reacts with the starting materials forming gaseous phases that can diffuse to the other end of the ampoule, normally kept at a lower temperature and called cold end. Crystals then deposit at the cold end and then the transport agent is released. In principle, the growth temperature can be significantly lower than the melting temperature of Bi$_2$Te$_3$, which makes possible the crystal growth of \MBT~at much lower temperatures. Up to now, vapor transport growth of \MBT~and related compounds has not been reported yet.

Motivated by purposely manipulating the concentration and distribution of lattice defects for more intrinsic or exotic quantum phenomena, in this work, we report the crystal growth of \MBT~and related compounds using vapor transport method. For the growth of \MBT~single crystals, I$_2$, MnI$_2$, MnCl$_2$, TeCl$_4$, or MoCl$_5$ are all valid transport agents. A small temperature gradient $<$20$^\circ$C is essential for the successful crystal growth of \MBT~and related compounds.  The lowest temperature that we managed to obtain \MBT~crystals is 500$^\circ$C when I$_2$ is used to assist the growth. Using I$_2$ as the transport agent, we also successfully grew single crystals of \MST, \MBST, and Sb-doped MnBi$_4$Te$_7$. Compared to flux grown crystals, vapor transported crystals tend to be Mn stoichiometric and the Sb-bearing compositions have more Mn site mixing. These differences demonstrate that the physical properties of \MBT~and related compounds can be engineered by using different materials synthesis routes.

\begin{table*}[tb]
\centering
\caption{Summary of the details of some selected growths mentioned in the text. Only the Sb content,$x$ in Mn$_2$Bi$_{2-x}$Sb$_x$Te$_5$, is listed for the starting composition of the last four batches with Sb doping. The tube furnace used for the growths is a single zone tube furnace with a thermocouple to monitor the temperature at the cold end of the ampoule. The Neel temperature, T$_N$, is defined as the temperature where a cusp was observed in the temperature dependence of magnetic susceptibility measured in a magnetic field of 1\,kOe. }
\begin{tabular}{c|c|c|c|c|c|c|c}
  \hline
  \hline
  Batch \#    &      starting composition    &      transport agent     &      T/time       & furnace&  EDS&  T$_N$/c & e/h($\times$10$^{19}$cm$^{-3}$)\\
 \hline
\#1164   &       Mn:Bi:Te=2:2:5   & I$_2$  &  585$^\circ$C, 10days   & box     &  Mn$_{0.94(2)}$Bi$_{2.07(1)}$Te$_{3.99(1)}$ & 25.0\,K & e, 3.1\\

\#1174B   &       Mn:Bi:Te=2:2:5   & TeCl$_4$  &  585$^\circ$C, 14days   & box     &  Mn$_{0.94(2)}$Bi$_{2.05(2)}$Te$_{4.01(3)}$ & 24.6\,K &e,7.3\\

\#1178   &       Mn:Bi:Te=2:2:5   & MnCl$_2$  &  585$^\circ$C, 21days   & box     &  Mn$_{1.01(2)}$Bi$_{1.99(2)}$Te$_{4.00(2)}$ & 25.2\,K&e, 5.0\\

\#1183B   &       Mn:Bi:Te=5:2:8   & I$_2$  &  550$^\circ$C, 14days   & tube     &  Mn$_{1.01(1)}$Bi$_{2.01(1)}$Te$_{3.98(1)}$ & 25.7\,K&e, 3.3\\

\#1198B   &       Mn:Bi:Te=2:2:5   & I$_2$  &  550$^\circ$C, 14days   & tube     &  Mn$_{0.99(1)}$Bi$_{2.00(1)}$Te$_{4.01(1)}$  & 25.7\,K&e, 1.2\\

\#1193A   &       Mn:Sb:Te=2:2:5   & I$_2$  &  590$^\circ$C, 14days   & box     &  Mn$_{1.15(2)}$Sb$_{1.95(1)}$Te$_{3.90(1)}$ & Tc=44\,K&p,110\\
\#1191A   &       $x$\,=\,0.65   & I$_2$  &  590$^\circ$C, 14days   & box     &  Mn$_{1.00(4)}$Bi$_{1.34(3)}$Sb$_{0.74(3)}$Te$_{3.92(3)}$ & 24.2\,K &--\\
\#1238   &       $x$\,=\,1.0   & I$_2$  &  585$^\circ$C, 14days   & box     &  Mn$_{0.97(4)}$Bi$_{1.04(1)}$Sb$_{1.05(2)}$Te$_{3.94(1)}$ & 23.5\,K&--\\
\#1193B   &       $x$\,=\,1.5   & I$_2$  &  590$^\circ$C, 14days   & box     &  Mn$_{1.01(3)}$Bi$_{0.50(3)}$Sb$_{1.53(1)}$Te$_{3.95(1)}$ & Tc=26\,K &--\\
\#1193C   &       $x$\,=\,1.0   & I$_2$  &  590$^\circ$C, 12days   & box     &  Mn$_{0.88(1)}$Bi$_{1.77(1)}$Sb$_{2.39(1)}$Te$_{6.96(1)}$ & 13\,K, 6\,K&p,37\\

 \hline
 \end{tabular}
\label{tab:Growths}
\end{table*}

\begin{figure*} \centering \includegraphics [width = 0.85\textwidth] {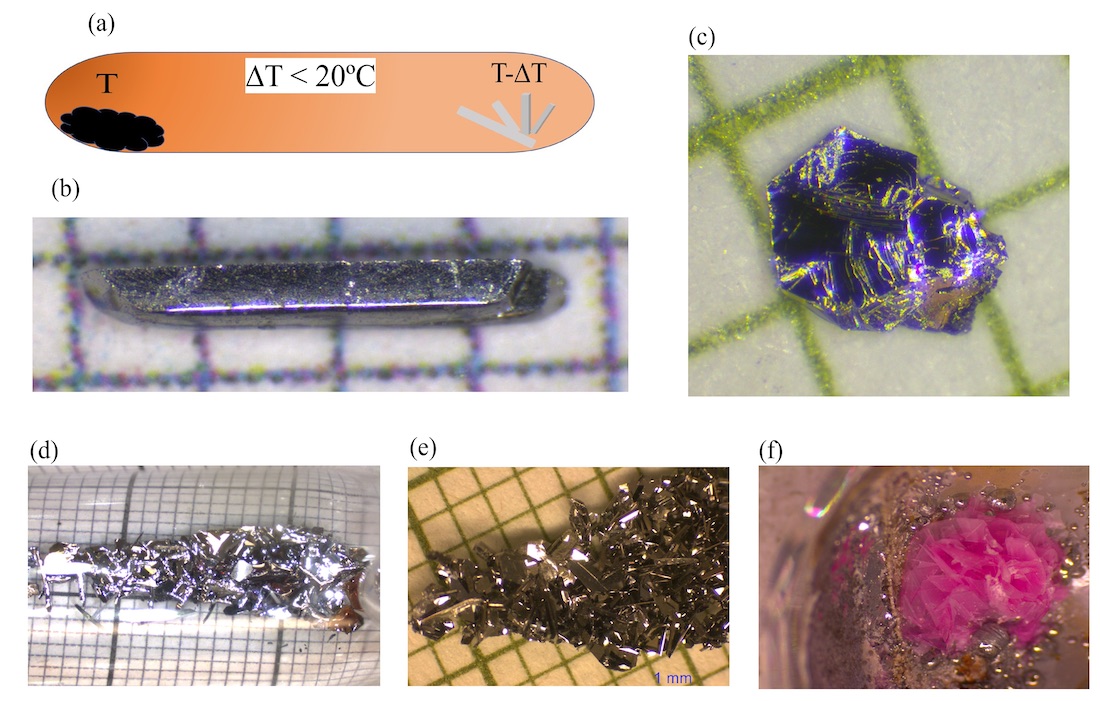}
\caption{(color online) (a) Schematic picture of the chemical vapor transport growth highlighting that a small temperature gradient $<$20$^\circ$C is essential for the successful growth of \MBT~and related compounds regardless the transport agents investigated in this work. (b) a rectangular-bar-shaped \MBT crystal from \#1164 on a millimeter grid. (c) a plate shaped \MBT crystal from \#1198B. (d) a mixture of rectangular-shaped crystals and plate-like crystals at the cold end of the growth ampoule. (e) cluster of plate-like crystals at the cold end of the growth ampoule. Crystals shown in (b-e) are grown using I$_2$ as the transport agent. (f) The droplets of Bi$_2$Te$_3$ melt and beautiful MnCl$_2$ crystals condensed at the cold end of growth ampoule for a vapor transport growth of \MBT~performed above 600$^\circ$C. }
\label{crystals-1}
\end{figure*}

\section{Experimental details}
Millimeter sized single crystals of \MBT, \MST, \MBST, and MnBi$_{4-x}$Sb$_x$Te$_7$ were grown by vapor transport method. Table I lists the detailed growth parameters of some selected growths. In a typical growth, 0.5\,g starting materials and 0.1\,g transport agent were sealed in a quartz tube of 12\,mm in inner diameter  and  4 inches in length. The growth was performed in a single zone tube furnace or a box furnace. When a tube furnace is used, an extra thermocouple is placed right next to the growth ampoule to monitor the temperature at the cold end. When a box furnace is used, the end pointing to the door is called the cold end assuming the area near the furnace door has a slightly lower temperature. The use of a box furnace is mainly due to the small temperature gradient needed for the crystallization of \MBT~and related compounds as presented later. After 2-3 weeks, single crystals are normally found at the cold end but occasionally along the tube, most likely due to the detailed temperature gradient. For the growth of \MBT, we obtained millimeter sized crystals using I$_2$, MnCl$_2$, TeCl$_4$, MnI$_2$, or MoCl$_5$ as the transport agent.

The elemental analysis was performed on freshly cleaved surfaces using a Hitachi TM-3000 tabletop electron microscope equipped with a Bruker Quantax 70 energy dispersive x-ray system. \textit{00l} reflections on cleaved surfaces were measured to confirm the phases  at room temperature using a PANalytical XPert Pro diffractometer with Cu-K$_{\alpha1}$ radiation. Magnetic properties were measured with a Quantum Design (QD) Magnetic Property Measurement System in the temperature range 2.0\,K$\leq$T$\leq$\,300\,K. The temperature and field dependent electrical resistivity data were collected using a QD Physical Property Measurement System (PPMS) in magnetic fields up to 140\,kOe. The scanning tunneling microscope (STM) and spectroscopy (STS) measurements were performed in an Omicron UHV-LT-STM with a base pressure 1$\times$10$^{-11}$mbar. Electrochemically etched tungsten tips were characterized on Au (111) surface.

\section{Results and discussion}

\subsection{Vapor transport growth of \MBT}
Halides were previously employed as the transport agent to grow MnBi$_2$Se$_4$\cite{nowka2017chemical}. We thus tested the growth of \MBT~using I$_2$,  MnI$_2$, MnCl$_2$, TeCl$_4$, or MoCl$_5$ as the transport agent. Considering the slow transportation of Mn\cite{nowka2017chemical},  we used Mn$_2$Bi$_2$Te$_5$ as the starting materials. All transport agents work well for the growth of \MBT~crystals. No difference was observed between growths using I$_2$ and MnI$_2$ as transport agent. However,  they both are more effective transport agents than those chlorides. As shown in Table I, growths using chlorides as transport agent require much longer growth time in order to obtain crystals of comparable size.

Figure\,\ref{crystals-1} (a) shows a schematic picture of the growth ampoule which highlights that a small temperature gradient $<$20$^\circ$C is necessary for the successful growth of \MBT~and related compounds. Once the temperature gradient is too large, Bi$_2$Te$_3$, normally doped with 1-4\% Mn, is always found at the cold end leaving some small MnTe crystals at the hot end of the growth ampoule. For growths performed inside of a tube furnace, plate-like crystals (see Fig.\,\ref{crystals-1}c) are always obtained regardless the type of transport agent, the temperature of the cold end, or the duration time. In contrast, for growths using a box furnace, both plate-like and rectangular-bar-shaped crystals (see Fig.\,\ref{crystals-1}b) are found inside of the growth ampoule. Figures\,\ref{crystals-1} (d, e) show the pictures of the cold end after the growth. There are more rectangular-bar-shaped crystals in Fig.\,\ref{crystals-1}d. Most \MBT~crystals stay at the cold end while occasionally crystals can be found in between the cold and hot ends along the tubes. This suggests that the temperature gradient inside of a box furnace is not that well defined as in a tube furnace. Despite the poorly defined temperature gradient, all growths of \MBT, \MBST, and \MST~are reproducible.

In addition to crystals of the desired composition, the following crystals can be occasionally found inside of the ampoule: MnTe$_2$, MnTe, Mn-doped Bi$_2$Te$_3$, and Mn$X_2$ ($X$=Cl, I). Mn$X_2$ crystals are transparent and moisture sensitive. The pyrite structured MnTe$_2$ grows like truncated cubes and can be easily identified by vision inspection under an optical microscope. The hexagonal structured MnTe crystals are also plate-like but rather brittle. Mn-doped Bi$_2$Te$_3$ and the targeted \MBT~and related compositions are quite alike and we normally distinguish them by using elemental ratio and (00l) reflections.

\begin{figure} \centering \includegraphics [width = 0.47\textwidth] {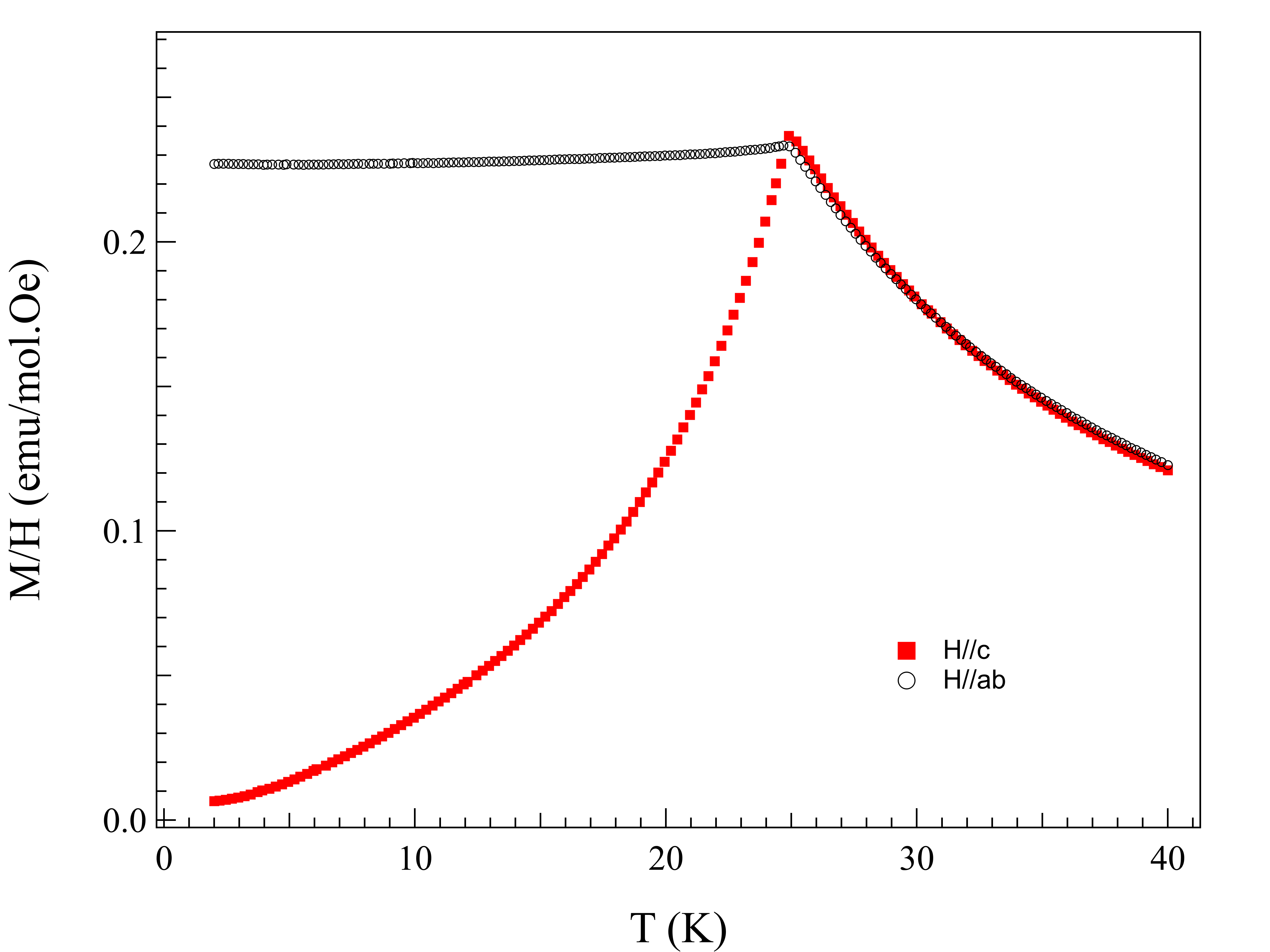}
\caption{(color online) Temperature dependence of magnetization measured in a field of 1kOe applied perpendicular (labelled as H//$ab$) and parallel to the crystallographic \textit{c}-axis.  This \MBT crystals (\#1164) was grown at 585$^\circ$C in a box furnace using I$_2$ as the transport agent.}
\label{chi}
\end{figure}

Vapor transport growth using MnCl$_2$ as the transport agent was specially investigated with great details aiming to obtain crystals good for the experimental realization of QAHE. The experimental realization of QAHE in \MBT~flakes is one of the most exciting results on this intrinsic magnetic topological insulator\cite{deng2020quantum}. But it is yet to be reproduced. In Ref.\,\citenum{deng2020quantum}, \MBT~crystals used for the device exhibiting QAHE were grown out of MnCl$_2$ flux. If the quality of the starting \MBT~crystals is the sole factor responsible for the absence of the spin flop transition and the observation of QAHE in Ref.\,\citenum{deng2020quantum}, crystal growth in the presence of MnCl$_2$ deserves special attention. Considering the vapor pressure of MnCl$_2$ and possible reaction between MnCl$_2$ and other starting materials at the growth temperatures used in Ref.\,\citenum{deng2020quantum}, one would expect that the vapor phases might play an important role in the crystal growth. We first determined the appropriate temperature range for the vapor transport growth of \MBT~with 0.1\,gram of MnCl$_2$ and 0.5\,gram of Mn$_2$Bi$_2$Te$_5$. All growths were performed in a box furnace using the small natural temperature gradient. Mn-doped Bi$_2$Te$_3$ crystals were obtained when the growth temperature is lower than 550$^\circ$C and Bi$_2$Te$_3$ liquids were observed at the cold end when the growth temperature is over 600$^\circ$C (see Fig.\,\ref{crystals-1}(f)). We then fixed the growth temperature at 585$^\circ$C but increased the amount of MnCl$_2$ in the starting materials from 0.1\,gram to 0.3\,gram and then to 0.5\,gram. With increasing amount of MnCl$_2$ in the starting materials, there are more transport pink colored MnCl$_2$ single crystals at the cold end after the crystal growth. However, Elemental analysis, magnetic measurement, and room temperature Hall measurements found all \MBT~crystals from these growths are comparable to each other. Thus the amount of MnCl$_2$ in the starting materials does not seem to affect the physical properties of \MBT~crystals obtained in the growths described above.

\begin{figure} \centering \includegraphics [width = 0.47\textwidth] {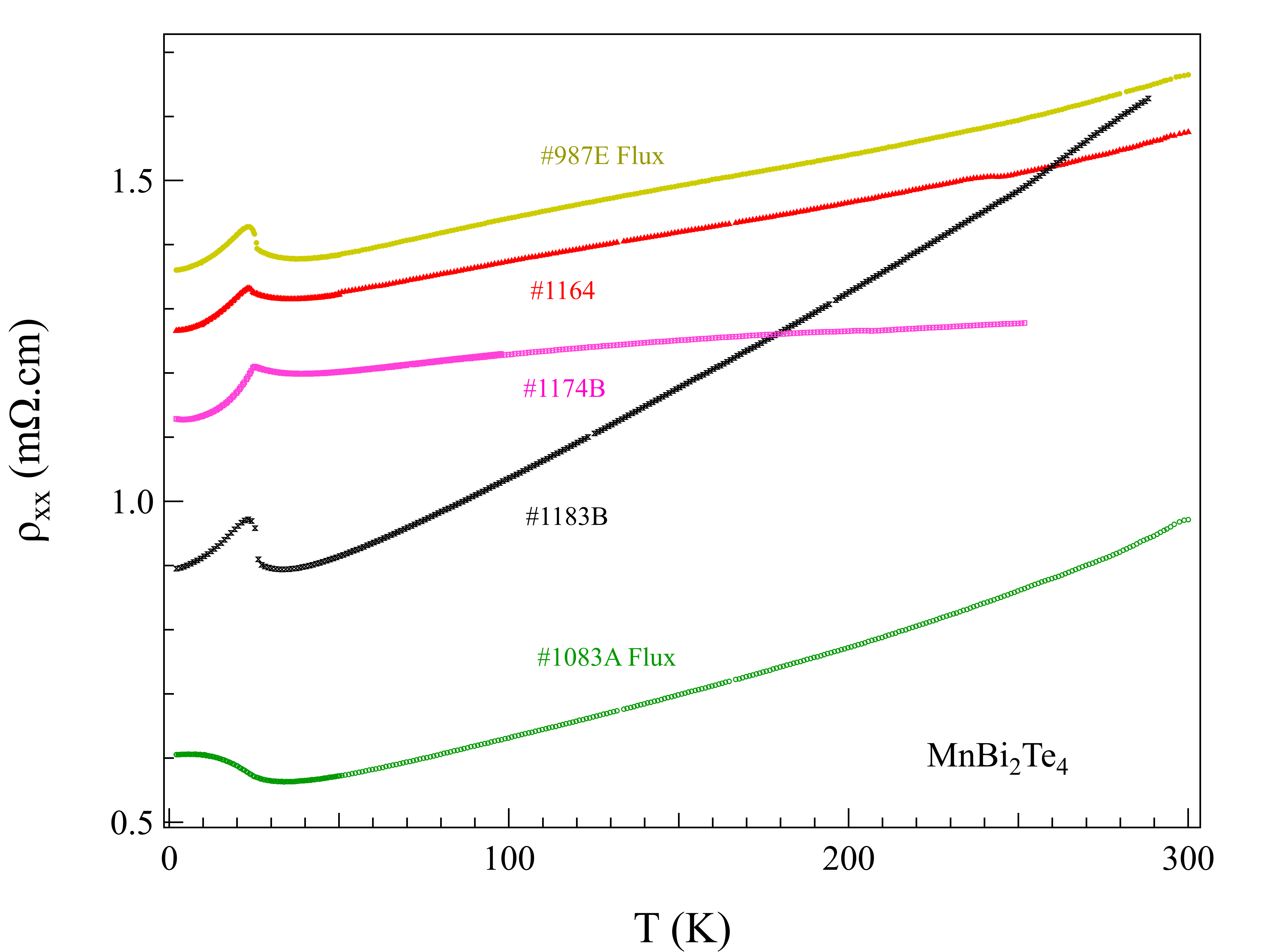}
\caption{(color online) Temperature dependence of in-plane electrical resistivity. The data for two batches of flux-grown crystals are also shown for comparison. The high field magnetization of \#1083A was reported in Ref.[\citenum{lai2021defect}]. Both batches of flux-grown crystals were used in the study of quantum phase transitions in thin flakes [Refs \citenum{ovchinnikov2021intertwined,cai2021electric}].}
\label{MBTRT}
\end{figure}

\begin{figure*} \centering \includegraphics [width = 0.95\textwidth] {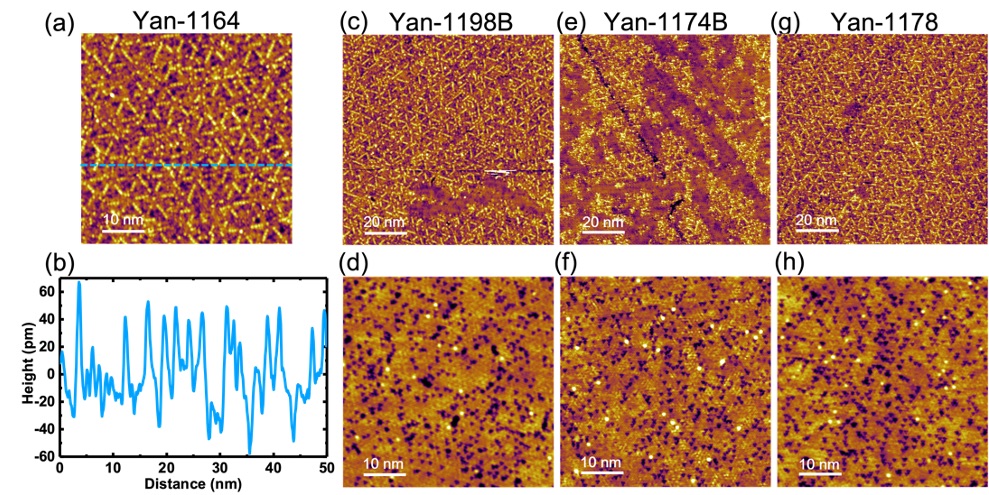}
\caption{(color online) (a) The topographic image of sample \#1164 surface after room-temperature cleaving. (b) The line profile along the blue dashed line in (a). (c) and (d) are respectively the topographic images of sample \#1198B surface after room-temperature and cold cleaving. (e) and (g) are respectively the topographic images of samples \#1174B and \#1178 surfaces after the sample is cold cleaved, warmed up to ~220K and inserted into the cold STM head. (f) and (h) are respectively the topographic images of samples \#1174B and \#1178 surfaces after cold cleaving. Tunneling set points of topographic images: (a) and (c) -1V, 30pA; (d) 1V, 100pA; (e) 1.5V, 30pA; (f) -0.8V, 100pA; (g) 1.5V, 30pA; (h) 1.5V, 20pA.}
\label{STM-1}
\end{figure*}

\begin{figure} \centering \includegraphics [width = 0.50\textwidth] {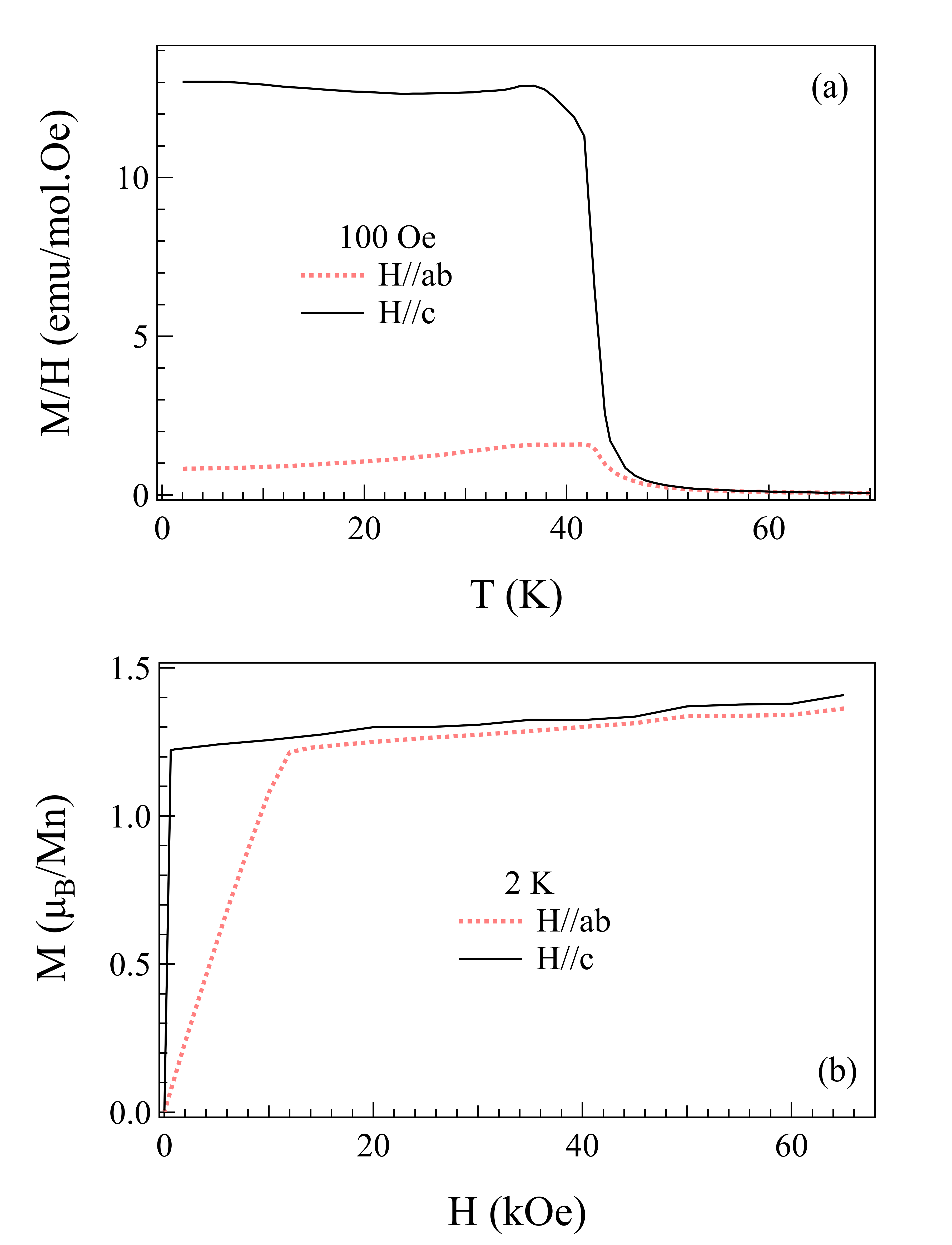}
\caption{(color online) Magnetic properties of \MST crystals. (a) Temperature dependence of magnetization with the magnetic field of 100\,Oe applied parallel (H$//$c) and perpendicular to the crystallographic $c$-axis.  (b) Field dependence of magnetization measured at 2\,K.}
\label{MSTMag-1}
\end{figure}

All \MBT~crystals we obtained in this study exhibit an antiferromagnetic order around T$_N$\,=\,25\,K. Figure\,\ref{chi} shows the temperature dependence of the magnetic susceptibility of one batch \#1164 as an example. The anisotropic temperature dependence agrees with previous report and suggests an antiferromagnetic order at T$_N$=25\,K. It should be mentioned that the magnetic data shown in Fig.\,\ref{chi} was collected without any cleaning after crystal growth. The data show no sign of the contamination of ferromagnetic Bi$_{2-x}$Mn$_x$Te$_3$, which has been occasionally observed when measuring the magnetic susceptibility of flux-grown \MBT~crystals\cite{yan2019crystal}.

We also measured the field dependence of magnetization at 2\,K in magnetic fields up to 120\,kOe in order to obtain the critical fields for the spin flop transition and the first magnetization plateau\citep{lai2021defect}. The correlation between T$_N$ and the critical field for spin flop transition is presented later. It is worth mentioning that the magnetization at 2\,K and 120\,kOe is around 3.8$\mu_B$/Mn for all the batches we measured. This suggests that all crystals we checked have similar amount of Mn/Bi site mixing, which is supported by the STM measurements presented later.

The temperature and field dependence of electrical resistivity was measured in the temperature range 2\,K$\leq$T$\leq$300\,K and in magnetic fields up to 120\,kOe. Figure\,\ref{MBTRT} shows the temperature dependence of the in-plane electrical resistivity of some vapor-transported crystals. The electrical resistivity data of two batches of flux grown crystals are also plotted for comparison. The temperature and field dependence agrees with previous reports.\cite{lee2018spin,zeugner2018chemical,otrokov2018prediction} From the Hall measurements, the electron density is in the range 1-8$\times$10$^{19}$cm$^{-3}$. The charge carrier concentration of some selected batches is listed in Table I.

\subsection{STM}

Figure\,\ref{STM-1}(a) shows the topographic image of the surface of sample \#1164 after room-temperature cleavage. Unlike the atomically flat surfaces seen in previous STM reports on our flux grown crystals\cite{yan2019crystal, liu2021site, huang2020native}, the surface here is covered with adatoms of height 60-80\,pm as inferred from Fig.\,\ref{STM-1}(b). Interestingly, the adatoms tend to align along the three crystallographic directions of the underlying triangular lattice, forming adatom chains. As shown in Fig.\, \ref{STM-1}(c), the room-temperature cleaved sample \#1198B show a similar surface. However, the atomically flat surface can be obtained after low temperature cleavage, as shown in Fig.\, \ref{STM-1}(d) for sample \#1198B. Measurements on other cold cleaved samples \#1174B and \#1178 shown in Fig. \, \ref{STM-1}(f) and (h) further confirm that cold cleavage exposes atomically flat and clean surface. To understand the origin of the adatoms appearing after room-temperature cleaving, we warmed up the cold cleaved samples to about 220 K and inserted them back in the STM head. As shown in Fig. \, \ref{STM-1}(e) and (g), adatoms appear on the surface and arrange in the similar pattern as that after room-temperature cleavage. 
It should be noted that different transport agents are used in the above batches mentioned. We also carefully monitored the $c$-lattice of these crystals and found no noticeable difference within experimental uncertainty. We thus speculate that these adatoms are the decomposition of transport agents (I or Cl ions) trapped on surfaces or grain boundaries during crystal growth. Immediately after cold cleavage, there is not enough thermal energy for them to diffuse so that we can observe clean surfaces. At elevated temperature these atoms could diffuse on the cleaved surface,  forming the pattern that we see in the STM images. This might need to be considered in ARPES measurements and crystal exfoliation for flakes.

From the atomically flat surface obtained after cold cleaving, the densities of the dominant defect Mn$_{Bi}$ antisites are (3.1$\pm$0.1)\% and (3.3$\pm$0.2)\% in sample \#1198B and \#1174B,  respectively. The density of Mn$_{Bi}$ in  \#1178  is  slightly higher at (4.0$\pm$0.2)\%.  This amount of Mn$_{Bi}$ defects is comparable to that in our flux grown crystals\cite{lai2021defect, huang2020native,yan2019crystal}.

\subsection{Vapor transport growth of \MST}
The same procedure can be employed to grow \MST~crystals. Figure\,\ref{MSTMag-1} shows the magnetic properties of \MST~crystals grown at 590$^\circ$C using I$_2$ as the transport agent. The temperature dependence of magnetization suggests a ferromagnetic inter-septuple-layer coupling and the magnetic ordering temperature, T$_c$, is about 44\,K. According to our previous work on \MST\cite{liu2021site}, Mn/Sb site mixing favors a ferromagnetic interlayer coupling and T$_c$ can be tuned from 24\,K to about 50\,K. The magnetic behavior shown in Figure\,\ref{MSTMag-1} signals a large amount of Mn/Sb antisite defects and few amount of Mn deficiency. The elemental analysis suggests a composition of Mn$_{1.15(2)}$Sb$_{1.95(1)}$Te$_{3.90(1)}$ which is Mn rich. The field dependence of magnetization at 2\,K (see Fig.\,\ref{MSTMag-1}(b)) can be employed to get a good estimate of the Mn/Sb antisite defects. The saturation moment is about 1.40$\mu_B$/Mn around 60\,kOe, which suggests about 72\% Mn site is taken by Mn assuming a local magnetic moment of 5$\mu_B$/Mn and a ferrimagnetic arrangement in each septuple layer\cite{liu2021site,lai2021defect}.

The electrical transport properties were also characterized by measuring the temperature dependence of electrical resistivity. A room temperature value of 1.6$\times$10$^{-3} \ohm.cm$ is similar to that of our flux grown crystals\cite{yan2019evolution, liu2021site}. Upon cooling, a metallic conducting behavior is observed and a quick drop of electrical resistivity was observed upon cooling through T$_c$. This temperature dependence is similar to that of ferromagnetic \MST~grown out of Sb$_2$Te$_3$ flux\cite{liu2021site}.  The Hall resistivity was measured at multiple temperatures in the temperature range 50~K~-~300~K. The Hall coefficient shows little temperature dependence below room temperature. With the assumption that a single band dominates the Hall signal, the coefficient at 250~K gives a carrier density of 1.1$\times$10$^{21}$cm$^{-3}$. Just like the flux grown crystals, the vapor transported \MST~crystals are also heavily hole doped.

\begin{figure*} \centering \includegraphics [width = 0.85\textwidth] {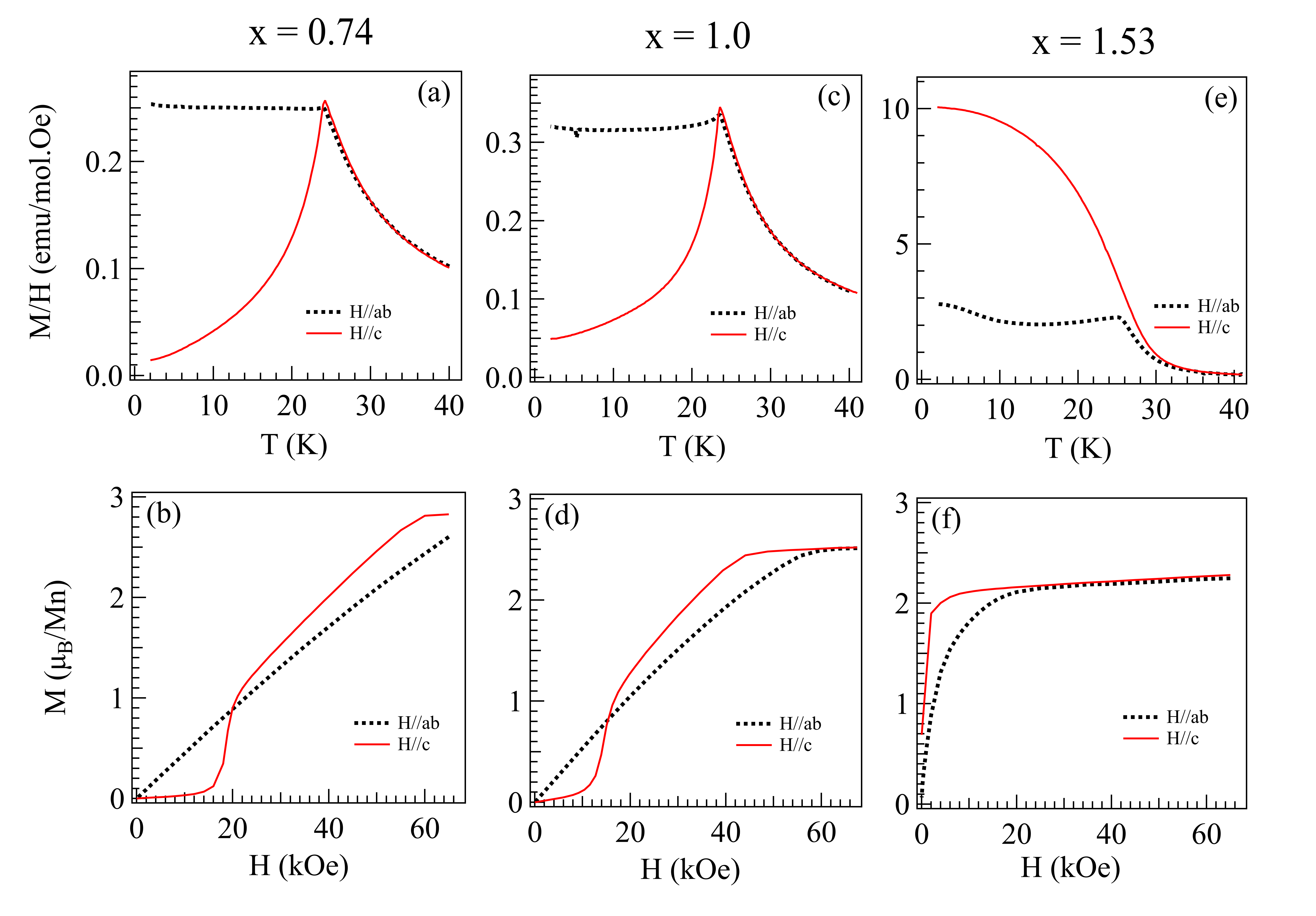}
\caption{(color online) Temperature and field dependence of magnetization for (a,b) $x$\,=\,0.74, (c, d) $x$\,=\,1.0, and (e, f) $x$\,=\,1.53 of \MBST crystals grown using I$_2$ as the transport agent.}
\label{MBSTMag-1}
\end{figure*}

Obviously, vapor transport growth can tune the chemistry of \MST~crystals in a wide range and the millimeter sized vapor transported crystals makes possible a careful investigation of the anisotropic properties. Compared to \MST~crystals grown out of Sb$_2$Te$_3$ flux at different temperatures, the I$_2$ transported crystals have more Mn ions residing on the Sb site which leads to a smaller saturation magnetization at low temperatures. We further tested the vapor transport growths at temperatures above 630$^\circ$C with the motivation of further increasing the concentration of Mn$_{Sb}$. However, only Mn-doped Sb$_2$Te$_3$ was obtained above 630$^\circ$C. Once the growth temperature is above the melting temperature of Sb$_2$Te$_3$, liquid Sb$_2$Te$_3$ droplets can be found at the cold end and no \MST~crystals could be obtained. We did not perform growths using other transport agents. However, it would be interesting to investigate whether other transport agents can affect the concentration and distribution of Mn defects and thus the magnetic and topological properties of \MST. In addition to the growth temperature and transport agent, the effect of the composition of the starting materials on the final nonstoichiometry of \MST~crystals also deserves a systematic investigation.

\subsection{Vapor transport growth of \MBST}

We further tested the growth of Sb-doped \MBT~crystals using I$_2$ as the transport agent. The focus of the current work is to explore routes to defect engineering of  the magnetic and topological properties by controlling the growth parameters rather than to perform a systematic study of the physical properties of \MBST. Figure\,\ref{MBSTMag-1} shows the magnetic properties of $x$\,=\,0.74, 1.0, and 1.53. The temperature dependence of magnetic susceptibility suggests an antiferromagnetic order for $x$\,=\,0.74 and 1.0. T$_N$, which is defined as the temperature where a cusp is observed in the temperature dependence of $\chi_c$, is 24.2\,K and 23.5\,K for $x$\,=\,0.74 and 1.0, respectively. T$_N$ for these two compositions agrees with that reported previously for flux grown crystals of similar compositions\cite{yan2019evolution}. The antiferromagnetic ground state is further confirmed by the field dependence of magnetization results shown in Fig.\,\ref{MBSTMag-1} (b,d) where a spin flop transition is observed. For $x$\,=\,1.53, the temperature and field dependence of magnetization suggests a ferromagnetic order below T$_c$=26\,K, similar to that observed in $x$=0.20 crystal grown by flux method\cite{chen2020ferromagnetism}.

Figs.\,\ref{MBSTMag-1} (b,d,f) show the field dependent magnetization at 2\,K. The absence of the metamagnetic transition further confirms the ferromagnetic order in $x$=1.53. With increasing Sb content, the magnetization at 65\,kOe decreases. This indicates that the amount of antisite Mn ions sitting at Bi/Sb site increases with increasing Sb content, which explains the ferromagnetism observed for $x$=1.53. Compared to the flux-grown crystals of similar compositions, the vapor transported \MBST~crystals have a smaller magnetization at 2\,K and 65\,kOe, suggesting that vapor transport growth using I$_2$ as transport agent leads to more site mixing between Mn and Bi/Sb. This has important consequences for $x$=0.74 and 1.0 on their critical magnetic fields, H$_{c1}$ for the spin flop transition and H$_{c2}$ for the magnetization plateau when the magnetic field is applied along the crystallographic$c$-axis.  For $x$\,=\,0.74, H$_{c1}$  is 19\,kOe and H$_{c2}$ is 60\,kOe. For $x$\,=\,1.0, H$_{c1}$  is 14\,kOe and H$_{c2}$ is 46\,kOe. For both compositions, both H$_{c1}$ and H$_{c2}$ are about 5-6\,kOe smaller than those in flux grown crystals with the same Sb content. When the magnetic field is applied perpendicular to the $c$-axis, the magnetization increases with field and saturates at H$_{ab}$ which is larger than H$_{c2}$. H$_{ab}$ decreases from above 70\,kOe for $x$\,=\,0.74 to 58\,kOe for $x$\,=\,1.0 with increasing Sb content. H$_{ab}$ is also smaller than that of flux grown crystals for these two compositions. All these observations support that vapor transported \MBST~crystals have more antisite defects than flux-grown crystals. This indicates that the detailed chemistry and physical properties of \MBST~crystals can be finely adjusted by using different growth techniques with different growth mechanism and kinetics.

It is worth mentioning that our EDS results suggest that Mn is near stoichiometric while Te tends to be deficient for all the \MBST~crystals we checked. As discussed later, this seems to be one difference between the vapor transport growth and the flux growth out of Bi$_2$Te$_3$.

\begin{figure} \centering \includegraphics [width = 0.47\textwidth] {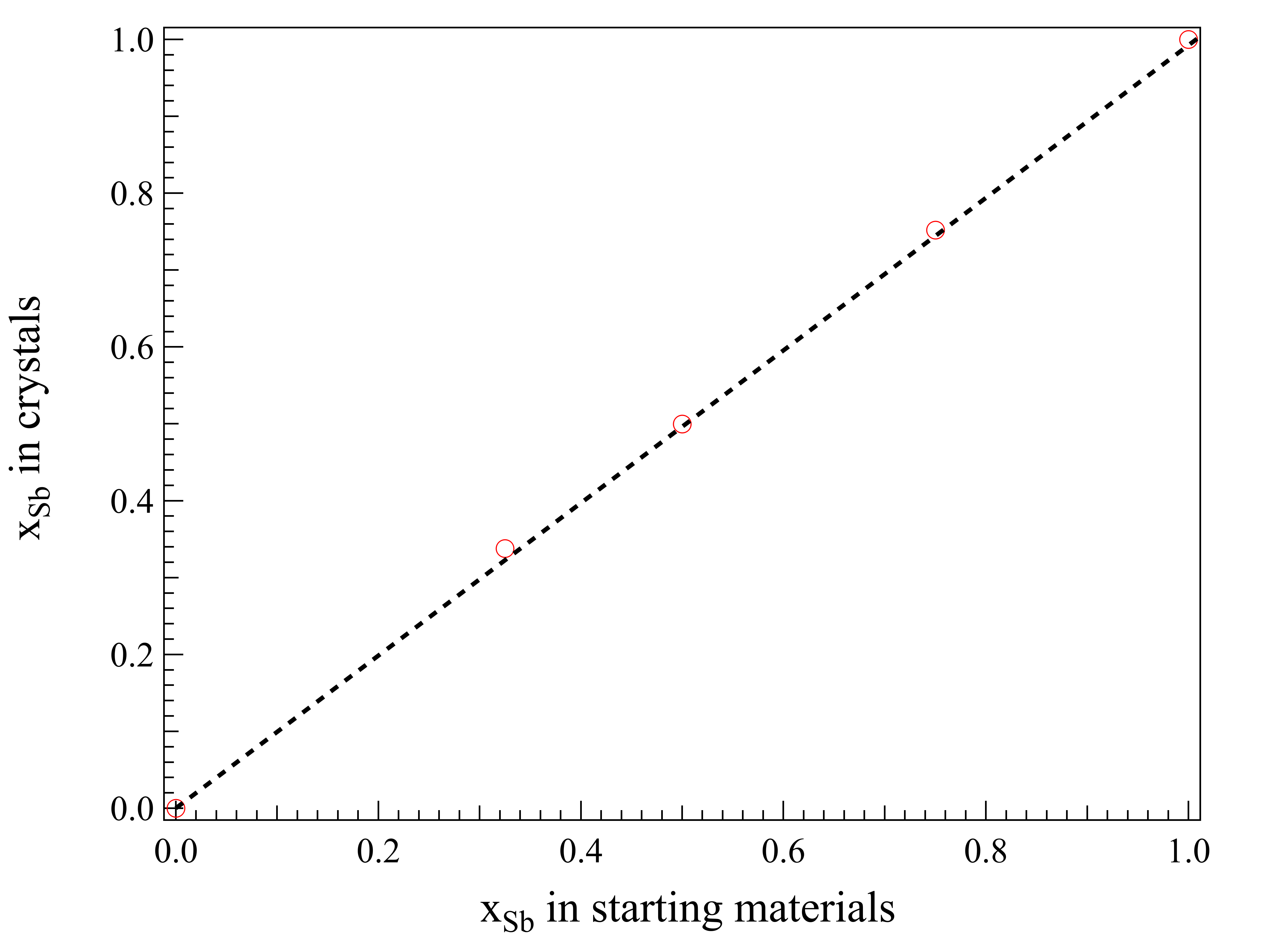}
\caption{(color online) Distribution of Bi/Sb in crystals and starting materials.}
\label{distribution-1}
\end{figure}

Figure\,\ref{distribution-1} shows the relation between the Sb concentration inside of the crystals and the nominal one in the starting materials. The amount of Sb inside of the crystals is close to that in the starting materials. This is in sharp contrast to that in growths out of Bi$_2$Te$_3$ flux where the Bi/Sb ratio can be quite off from the starting compositions due to the distribution coefficient of (Bi) and (Sb) in the flux\cite{yan2019evolution}.  The consistency between the composition of the starting materials and single crystals enables an easier control of the composition and easier targeted growths in future vapor transport growths of \MBST~crystals.

\subsection{Vapor transport growth of MnBi$_{4-x}$Sb$_x$Te$_7$}

\begin{figure} \centering \includegraphics [width = 0.47\textwidth] {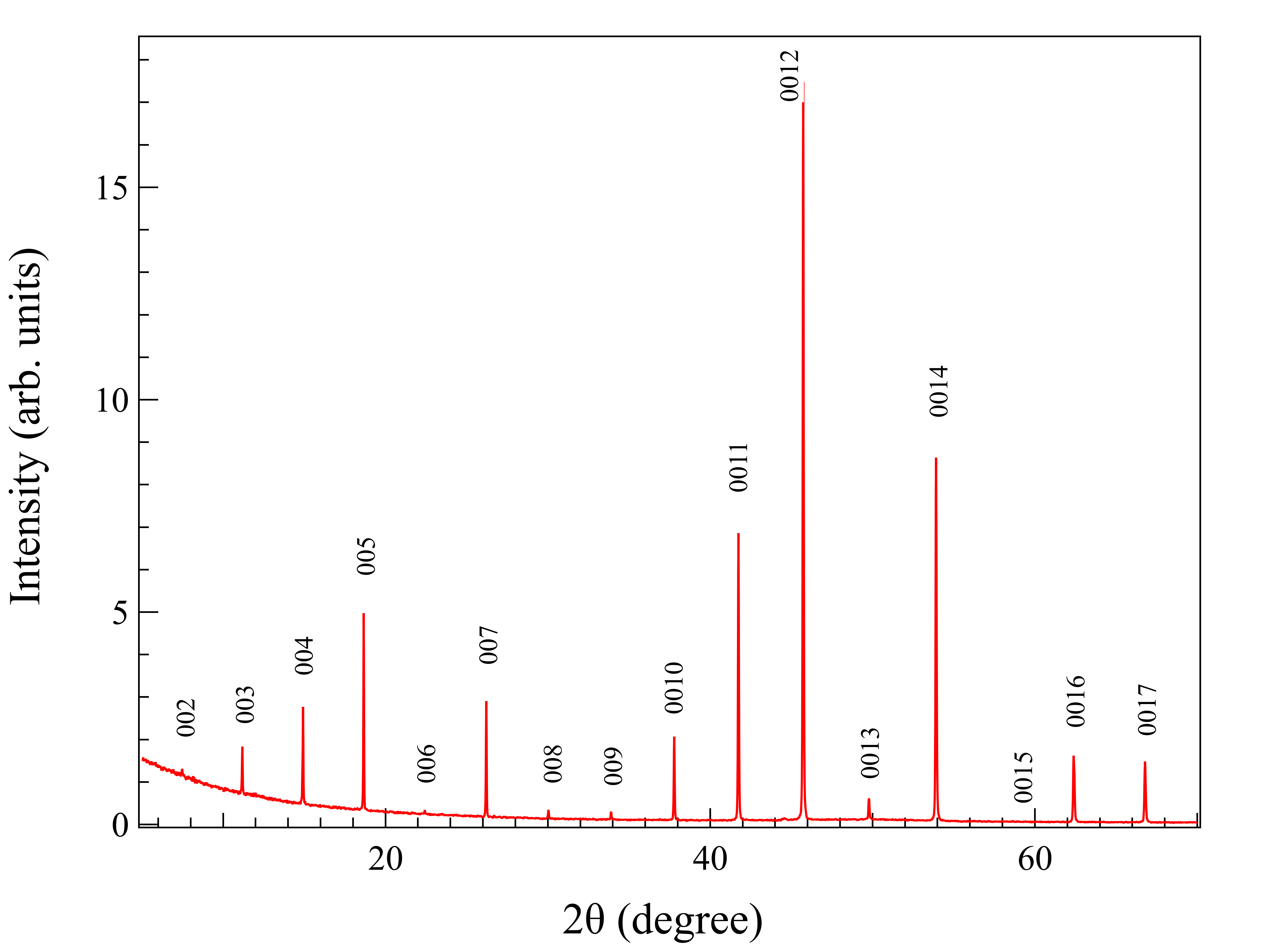}
\caption{(color online) ($00l$) reflections of MnBi$_{4-x}$Sb$_x$Te$_7$ suggesting the c-axis is perpendicular to the plane of the plate.}
\label{XRD-1}
\end{figure}

\begin{figure} \centering \includegraphics [width = 0.47\textwidth] {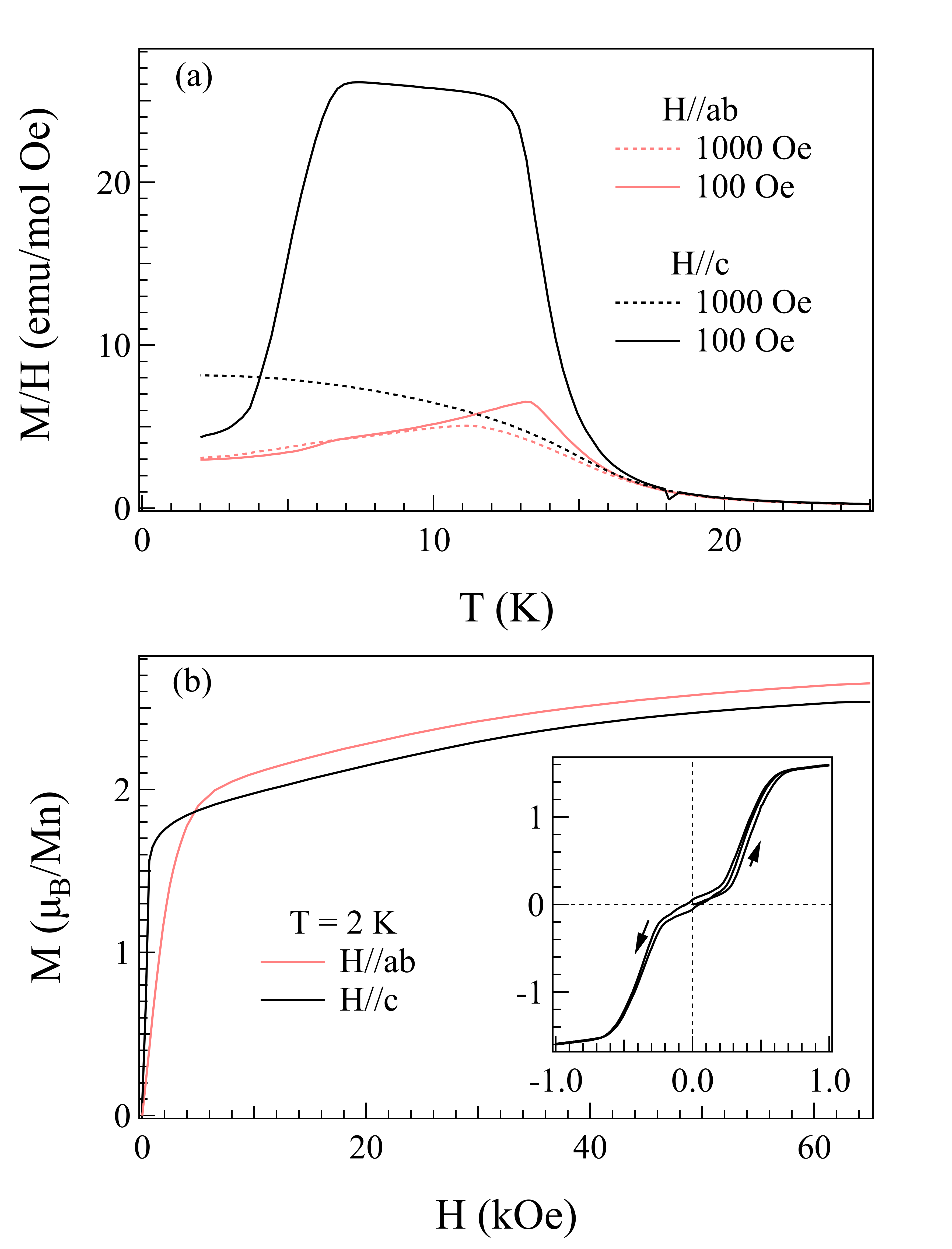}
\caption{(color online) (a) The temperature dependence of magnetization. (b) The field dependence of magnetization at 2\,K. Inset shows the spin flip transition around 300\,Oe when H//$c$.}
\label{MTH147}
\end{figure}

In one test growth (batch \#1193C) of \MBST~using I$_2$ as the transport agent, MnBi$_{4-x}$Sb$_x$Te$_7$ crystals were obtained instead. The growth details are listed in Table 1. Elemental analysis suggests a composition of Mn$_{0.88(2)}$Bi$_{1.76(1)}$Sb$_{2.39(1)}$Te$_{6.96(2)}$.
Figure\,\ref{XRD-1} presents the $(00l)$ reflections collected using x-ray diffraction on the surface of the plate-like single crystals. These $(00l)$ reflections can be indexed with the structure of MnBi$_4$Te$_7$  and the $c$ lattice is 23.794(3)$\AA$, close to that of MnBi$_4$Te$_7$ reported previously\cite{souchay2019layered,hu2020van,yan2020type}. Figure\,\ref{MTH147} shows the temperature dependence of the magnetic susceptibility measured in magnetic fields applied perpendicular (labelled as H//$ab$) and parallel (H//$c$) to the crystallographic \textit{c}-axis. All measurements were carried out in a field cooled mode. The anisotropic magnetic susceptibility data collected in an applied magnetic field of 1\,kOe  suggest a ferromagnetic interlayer coupling with the moments aligned along the crystallographic $c$-axis. The magnetic ordering temperature is about 14\,K, which is close to that reported for Sb-doped MnBi$_4$Te$_7$\cite{hu2021tuning, chen2021coexistence}. When a magnetic field of 100\,Oe is applied along $c$-axis during the measurements, a sharp drop of magnetization is observed upon cooling below about 6\,K. A much weaker anomaly was observed at 6\,K when this small field is applied perpendicular to the $c$-axis. This suggests the appearance of antiferromagnetic alignment of magnetic moments. The difference between the magnetization data collected in 100\,Oe and 1\,kOe suggests that a small field near 1\,kOe is enough to polarize the antialigned spins. It should be noted that the mexican-hat-like temperature dependence of magnetization is not observed in flux grown MnBi$_{4-x}$Sb$_x$Te$_7$ crystals\cite{hu2021tuning, chen2021coexistence}.  The magnetic susceptibility curves (not shown) measured with H//ab and H//c in the temperature range 50$\leq$T$\leq$300\,K can be described by the Curie-Weiss law, $\chi=C/(T-\theta$), where C is the Curie constant and $\theta$ is the Weiss temperature. The Curie-Weiss fitting of the high temperature susceptibility data obtains a Weiss temperature of 6\,K and an effective moment of 4.95\,$\mu_B$/Mn.

The mexican-hat-like temperature dependence of magnetization shown in Fig.\,\ref{MTH147}(a) suggests  a ferromagnetic order around 14\,K and then a transition upon cooling to an antiferromagnetic order around 6\,K.  The sharp drop of magnetization around 6\,K is more likely due to the change of the inter-septuple-layer coupling from ferromagnetic to antiferromagnetic other than a sudden alignment of Mn$_{Bi}$ moments antiparallel to Mn$_{Mn}$. First, Mn$_{Bi}$ in the septuple layers are strongly antiferromagnetically coupled to Mn$_{Mn}$ and a large magnetic field is needed to change their alignment\citep{lai2021defect}. Second, it is not clear whether and how the Mn$_{Bi}$ ions in the quintuple layers of MnBi$_4$Te$_7$ order magnetically. A recent study\citep{hu2021tuning} on Sb-doped MnBi$_4$Te$_7$ proposed the absence of any long range magnetic order of Mn$_{Bi}$ ions in the quintuple layers but it is yet to be confirmed experimentally. It is reasonable to speculate that the state of Mn$_{Bi}$ ions in the quintuple layers depends on the concentration and distribution of magnetic defects. However, considering the small population of Mn$_{Bi}$ ions in the quintuple layers, one would not expect such a large drop of temperature dependence of magnetization around 6\,K. This is supported by the large magnetization change at the spin flip transition presented in Fig.\,\ref{MTH147}(b). Finally, previous studies show that the energy difference between antiferromagnetic and ferromagnetic inter-septuple-layer coupling is small in \MBT~and related compounds and the delicate magnetic ground state is fragile and sensitive to the inter-septuple-layer spacing or lattice defects. The  mexican-hat-like temperature dependence of magnetization shown in Fig.\,\ref{MTH147} (a)  should result from a temperature driven change of the magnetic ground state.

With an antiferromagnetic inter-septuple-layer coupling below 6\,K, one would expect a metamagnetic transition in applied magnetic fields. Figure\,\ref{MTH147}(b) shows the field dependence of magnetization at 2\,K in fields up to 70\,kOe applied along or perpendicular to the crystallographic \textit{c} axis. The inset highlights the spin flip transition at a critical field about 300\,Oe, which disappears when measured above 6\,K. This confirms that our sample shows a thermal induced change of magnetic order. The magnetization measured at 2\,K and 65\,kOe is about 2.6$\mu_B$/Mn, which is much smaller than the expected value and signals that there are significant amount of antisite Mn ions antiparalllel to those Mn ions in the main Mn layer. In \MBT~and \MST, the antiferromagnetic coupling between Mn$_{Mn}$ and Mn$_{Bi/Sb}$ in the same septuple layer is strong and a large magnetic field near 50\,T is needed to polarize all Mn moments. This strong coupling is also expected in the septuple layers in MnBi$_4$Te$_7$. With the assumption that Mn ions at Bi site in the quintuple layers are already polarized in a magnetic field of 70\,kOe, the magnetization of about 2.6$\mu_B$/Mn suggests that 20\% of the Mn ions stay at the Bi sites of the setuple layers. Similar to that observed in the MnBi$_4$Te$_7$, both H//$ab$ and H//$c$ magnetization curves show an increase of about 10\% in the field range 10-40\,kOe. Previously, we speculated that this 10\% increases comes from the polarization of the moment on those Mn ions sitting at Bi site in the quintuple layers\cite{yan2020type}. If these Mn ions do not exhibit a spontaneous magnetic order, the field required for this 10\% magnetization change should signal the exchange field from the neighbouring septuple layers.

\begin{figure} \centering \includegraphics [width = 0.45\textwidth] {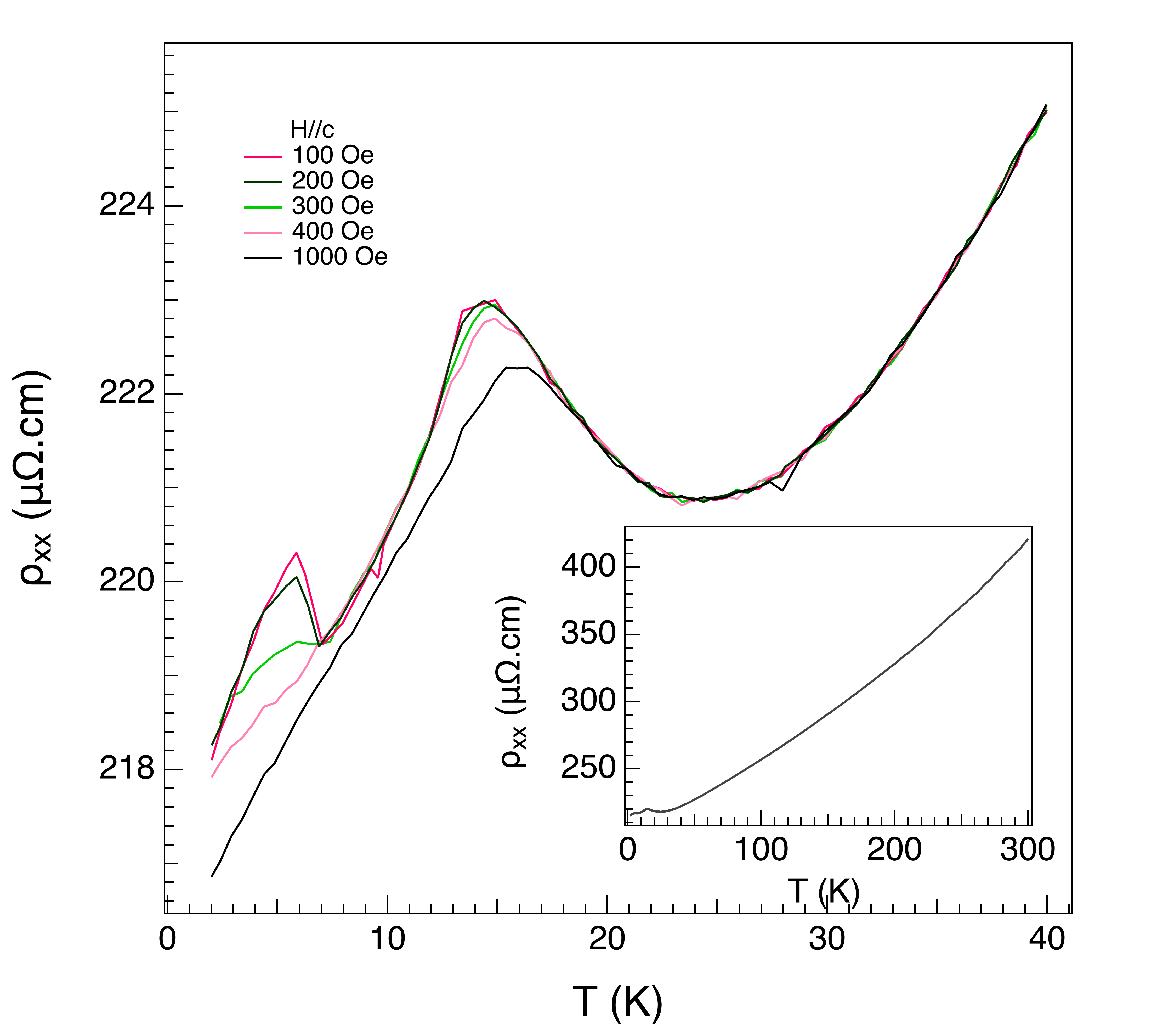}
\caption{(color online) Temperature dependence of in-plane electrical resistivity of MnBi$_{4-x}$Sb$_x$Te$_7$ in small magnetic fields applied along the $c$-axis. The inset shows the data in the temperature range 2-300\,K in zero magnetic field. }
\label{RT147-1}
\end{figure}

Figure\,\ref{RT147-1} shows the temperature dependence of the in-plane electrical resistivity with the electrical current flowing in the $ab$-plane. As shown in the inset, a metallic conducting behavior was observed and there are two cusps at low temperatures where the mexican hat like magnetic susceptibility was observed. When a small magnetic field is applied along the $c$-axis, the cusp centering around 14\,K shifts to higher temperatures,  while the one around 6\,K is suppressed to lower temperatures and disappears above 400\,Oe. This interesting field dependence signals that the transition around 6\,K is fragile and is consistent with the spin flip transition shown in inset of Fig.\,\ref{MTH147}(b).

We would like to mention that this is the only batch of MnBi$_{4-x}$Sb$_x$Te$_7$ single crystals that we obtained in over 10 similar growths. It is likely that the temperature gradient along the growth ampoule happens to be in the appropriate range for the crystallization of MnBi$_{4-x}$Sb$_x$Te$_7$. This is similar to the flux growth of MnBi$_4$Te$_7$ crystals out of Bi$_2$Te$_3$ flux and it is understandable considering all MnTe.$m$Bi$_2$Te$_3$ phases exist in a narrow temperature range in the pseudobinary MnTe-Bi$_2$Te$_3$  system\cite{aliev2019novel}.  Although it is unfortunate that we failed to identify the exact growth conditions and to understand the growth mechanism, this only successful growth demonstrates that more complex MnTe.$m$Bi$_2$Te$_3$ compounds can be obtained by the chemical vapor transport growth. Compared to MnBi$_4$Te$_7$ crystals grown out of Bi$_2$Te$_3$ flux, our MnBi$_{4-x}$Sb$_x$Te$_7$ crystals is less Mn deficient. The Mn deficiency in MnTe.$m$Bi$_2$Te$_3$ increasing with $m$ driven by the increasing lattice mismatch between the MnTe and the Bi$_2$Te$_3$ in the nature heterostructure\citep{du2021tuning}.  The large Mn deficiency might prevent the experimental observation of the theoretically predicted quantum phenomena in MnTe.$m$Bi$_2$Te$_3$ with $m>$1. Chemical vapor transport growth reported in this work provides a different approach for the materials synthesis and a careful control of the growth process might eventually lead to a good control of magnetic defects (Mn$_{Bi}$ and magnetic vacancy at Mn site) thus for the intrinsic properties and maybe the theoretically predicted topological phenomena.

\subsection{Other variants}
$n$MnTe.Bi$_2$Te$_3$, with more than one MnTe sheet inserted into Bi$_2$Te$_3$, might show complex magnetism and host novel topological properties\cite{zhang2020large,li2020intrinsic, lei2020magnetized}. A recent study reports the successful growth of $n$=2 member, Mn$_2$Bi$_2$Te$_5$, by flux method\cite{cao2021growth}. It is interesting to note that we found only MnTe.\textit{m}Bi$_2$Te$_3$ crystals even though the starting materials are $n$MnTe.Bi$_2$Te$_3$ with $n$ up to 5. This indicates the current growth parameters are not ideal for the growth of $n$MnTe.Bi$_2$Te$_3$  single crystals. As presented earlier, once the growth temperature is below 500$^\circ$C or above 600$^\circ$C, only Bi$_2$Te$_3$ doped with 1-4\% Mn was obtained. The growth below 500$^\circ$C is limited by the effective transport of Mn. Thus, growths using other transport agents which can transport Mn more effectively should be investigated which might help obtain $n$MnTe.Bi$_2$Te$_3$ compounds.

\begin{figure} \centering \includegraphics [width = 0.45\textwidth] {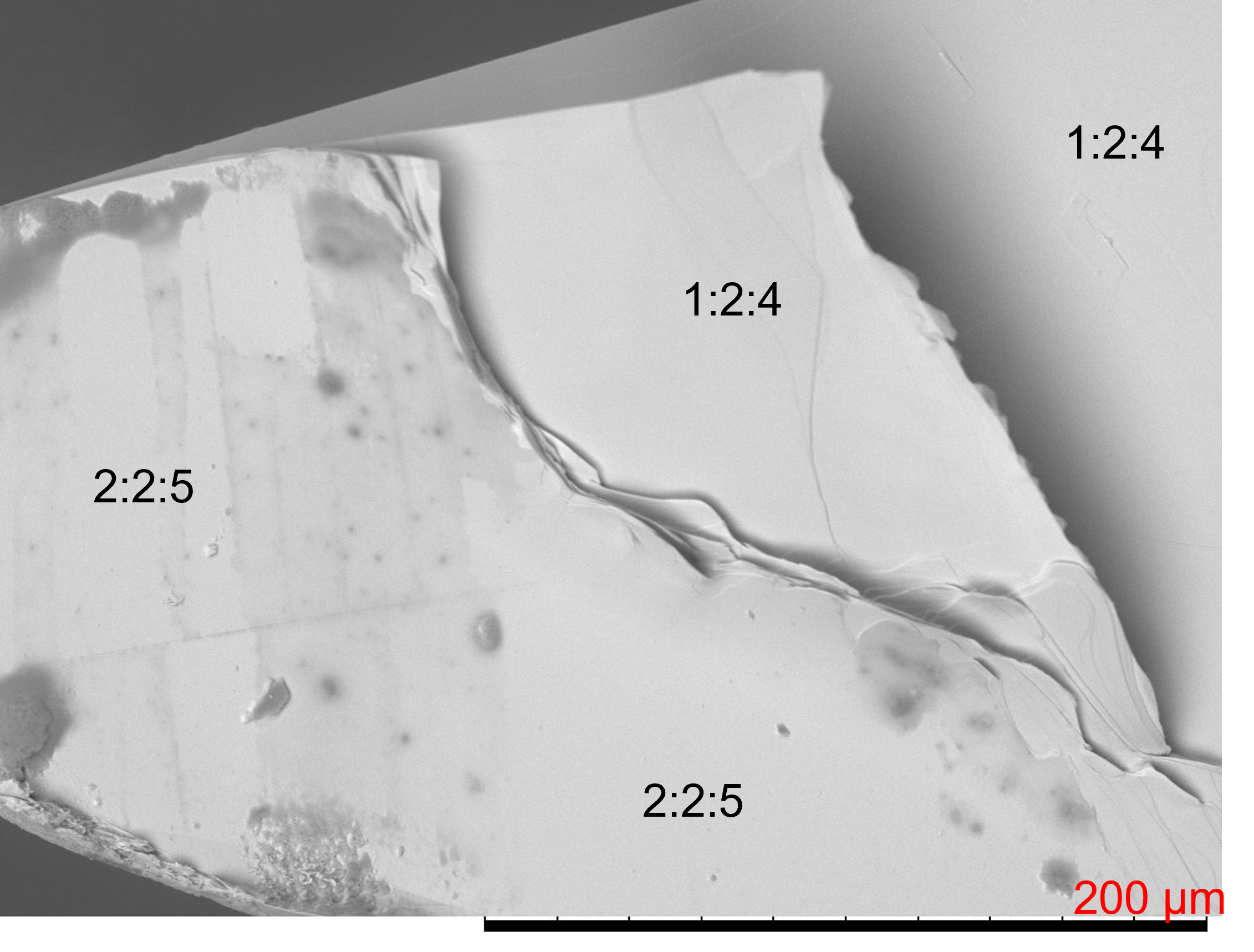}
\caption{(color online) SEM image of the cleaved surface of one crystal from batch \#1198B and the elemental analysis of different regions. }
\label{Yan1198B-1}
\end{figure}

It is noteworthy that our elemental analysis did observe regions with the elemental ratio consistent with Mn$_2$Bi$_2$Te$_5$ in one \MBT~crystal. Figure\,\ref{Yan1198B-1} shows the SEM image of one cleaved surface of one piece of \MBT~crystal from \#1198B. While we observed Mn:Bi:Te=1:2:4 in most positions on this surface, the top flake was found to have Mn:Bi:Te=1.96:2.14:4.90. We carefully looked at the regions nearby but failed to find any sign of MnTe, therefor, the observed elemental ratio of Mn:Bi:Te=1.96:2.14:4.90 doesn't seem to be from a mixture of \MBT~and MnTe.  It is unfortunate that we failed to isolate enough pieces  with that elemental ratio for further characterization. More growths are in progress to identify the appropriate growth parameters for the crystallization of Mn$_2$Bi$_2$Te$_5$. However, the presence of that Mn$_2$Bi$_2$Te$_5$-like flake indicates that (1) vapor transport growth might be employed to grow Mn$_2$Bi$_2$Te$_5$, (2) some strain field or nonequilibrium conditions might help the crystallization of Mn$_2$Bi$_2$Te$_5$, and  (3) the intergrowth of Mn$_2$Bi$_2$Te$_5$ and \MBT~can occur and should be considered in future growths and characterization of Mn$_2$Bi$_2$Te$_5$.

Introducing Cr or V into Mn or Bi site might be able to tune the magnetism of \MBT. For example, Mn$_{1-x}$V$_x$Bi$_2$Te$_4$  is proposed to show a ferromagnetic inter-septuple-layer coupling which is preferred for the observation of QAHE\cite{hou2021alloying}. Partial substitution of Bi by Cr or V might be another way toward a ferromagnetic interlayer coupling since these magnetic defects are expected to favor a ferromagnetic inter-septuple-layer coupling. We thus performed vapor transport growth starting with premelted MnBi$_{1.9}$\textit{M}$_{0.1}$Te$_4$ ($M$=Cr, V) using I$_2$ as transport agent at 585$^\circ$C in a box furnace. In all test growths, the obtained single crystals show a T$_N$ of 25\,K. Elemental analysis does not suggest any successful substitution. One weak anomaly was observed around 200\,K in the temperature dependence of magnetic susceptibility for the crystals grown out of MnBi$_{1.9}$Cr$_{0.1}$Te$_4$ premelt. This suggests the presence of Cr$_2$Te$_3$. We further increased the amount of $M$Te in the starting materials and tried growths starting with equal mass of Mn$_2$Bi$_2$Te$_5$ and $M$Te. Elemental analysis and magnetic measurements do not suggest any incorporation of Cr or V in \MBT~crystals. Vapor transport growths using $MX_3$ ($M$=Cr, V, $X$=Cl, Br, I)  as transport agent might deserve some effort as all $MX_3$ have a reasonably high  vapor pressure around 600$^\circ$C.

\subsection{Vapor transport vs flux growth}

Both chemical vapor transport and flux growths can grow single crystals of \MBT~and related compounds of similar quality. They both can be employed to grow crystals for various measurements. The vapor transport growth developed in this work expands our capability of fine tuning of the nonstoichiometry, concentration and distribution of lattice defects, and hence the magnetic and topological properties of these fascinating compounds. Understanding the difference between vapor transported and flux grown crystals and also that between these two growth techniques will help us select the appropriate crystals or growth technique that could best address the physics questions or meet the scientific needs.

(1) Vapor transported crystals tend to be more Mn stoichiometric. As listed in Table I, once the temperature gradient is small enough, all \MBT~and \MBST~crystals are Mn stoichiometric. In our vapor transported \MBT~crystals, our STM results shown in Fig.\,\ref{STM-1} suggest that there are about 3-4\% Mn$_{Bi}$, comparable to what we observed in flux grown \MBT~crystals\cite{yan2019crystal, huang2020native}. This is also supported by magnetic measurements which show the vapor transported and flux-grown crystals have similar magnetic ordering temperature, saturation moment, critical magnetic fields for spin flop transition and moment saturation at low temperatures. The difference in Mn nonstoichiometry in vapor transported and flux-grown \MBT~crystals results from the different growth mechanisms. In crystal growth using Bi$_2$Te$_3$ as flux, Mn content in \MBT~crystals depends on the distribution coefficient of (Mn) between the flux and the crystal, which normally depends on both temperature and composition. Considering the growth temperature is close to the melting temperature of Bi$_2$Te$_3$, the diffusivity of (Mn) in Bi$_2$Te$_3$ flux is expected to be small, which can further lead to Mn deficiency even though a rather slow cooling rate has been employed. The Mn content in flux-grown crystals can be purposely controlled by using carefully selected growth parameters such as the composition of starting materials and temperature profiles. For vapor transported growths, the effective transportation of Mn seems to control the nonstoichiometry. As shown in Table I, Mn-deficient \MBT~crystals were obtained in several early batches for which a slightly larger temperature gradient has been used. We also tried to add more MnTe in the starting materials and this does not seem to change the total Mn content in the resulting crystals. A careful monitoring of the correlation between crystal  chemistry and growth parameters suggests the temperature gradient plays a key role in determining the Mn nonstoichiometry when we fix the growth temperature around 580$^\circ$C. If this understanding is right, an even smaller temperature gradient may facilitate the growth of Mn$_2$Bi$_2$Te$_5$.

(2) Vapor transported Sb-bearing crystals tend to have more antisite Mn ions.  The vapor transported \MST~crystals are Mn-rich (Table I) and have a ferromagnetic ordering temperature of about 44\,K, consistent with previous studies that Mn$_{Sb}$ favors a ferromagentic interlayer coupling and the magnetic ordering temperature can be fine tuned in a wide temperature range 20-50\,K\cite{liu2021site}. The field dependence of magnetization curves shown in Fig.\,\ref{MBSTMag-1} also suggest that  Sb-doped \MBT~crystals grown by vapor transport have more antisite magnetic defects. This can have important implications for the fine tuning of the magnetic and topological properties of Bi-rich \MBST. Previous studies\citep{yan2019evolution, chen2019intrinsic} have demonstrated that Sb-doping can effectively tune the physical properties of \MBT. Vapor transport growth makes possible further tuning of the properties in a sample with a fixed Sb content  via controlling the distribution of Mn at different crystallographic sites. 

(3) Vapor transported crystals are smaller in dimension than flux grown crystals.  This is especially true when chlorides are used as the transport agent. Most crystals are in sub-millimeter size and less than 2\,mg/piece, which are at least one order of magnitude smaller than the flux-grown crystals. Thus flux growth is more appropriate when large crystals or a large amount of crystals are needed for some special measurements such as inelastic neutron scattering. This doesn't mean any sacrifice of the crystal quality because large  stoichiometric \MBT~crystals can be grown out of Bi$_2$Te$_3$ flux with carefully controlled growth parameters.

(4) Vapor transported \MBT~crystals are free of contamination by magnetic Mn-doped Bi$_2$Te$_3$. This is well illustrated by the magnetic susceptibility results shown in Fig.\ref{chi}. While for flux grown crystals, a careful cleaning is normally needed to remove residual flux on the surface\cite{yan2019crystal}. However, as shown in Fig.\,\ref{STM-1}, the residual transport agent can contaminate the cleaved surface of vapor transported crystals.

(5) Vapor transport technique can have an easier control of the Sb content in \MBST~crystals than the flux growth. This is demonstrated by the correspondence of Bi/Sb in crystals and starting materials shown in Fig.\,\ref{distribution-1}. As we have not been able to identify the right growth conditions for MnBi$_{4-x}$Sb$_x$Te$_7$, it is not clear whether Bi/Sb ratio in crystals would be the same as that in the starting materials for MnBi$_{4-x}$Sb$_x$Te$_7$. The only data point (\#1193C) that we have signals a more complex relation.

(6) Vapor transport growths of \MBT~crystals were performed in a wider temperature range 500$^\circ$C-590$^\circ$C. The highest possible growth temperature of our vapor transport growths is limited by the melting temperature of Bi$_2$Te$_3$, and the lowest temperature by the effective transport of all constituents. While for flux growths, the highest and lowest growth temperature is determined by the peritectic decomposition temperature of \MBT~and the melting temperature of Bi$_2$Te$_3$, respectively.

\begin{figure} \centering \includegraphics [width = 0.5\textwidth] {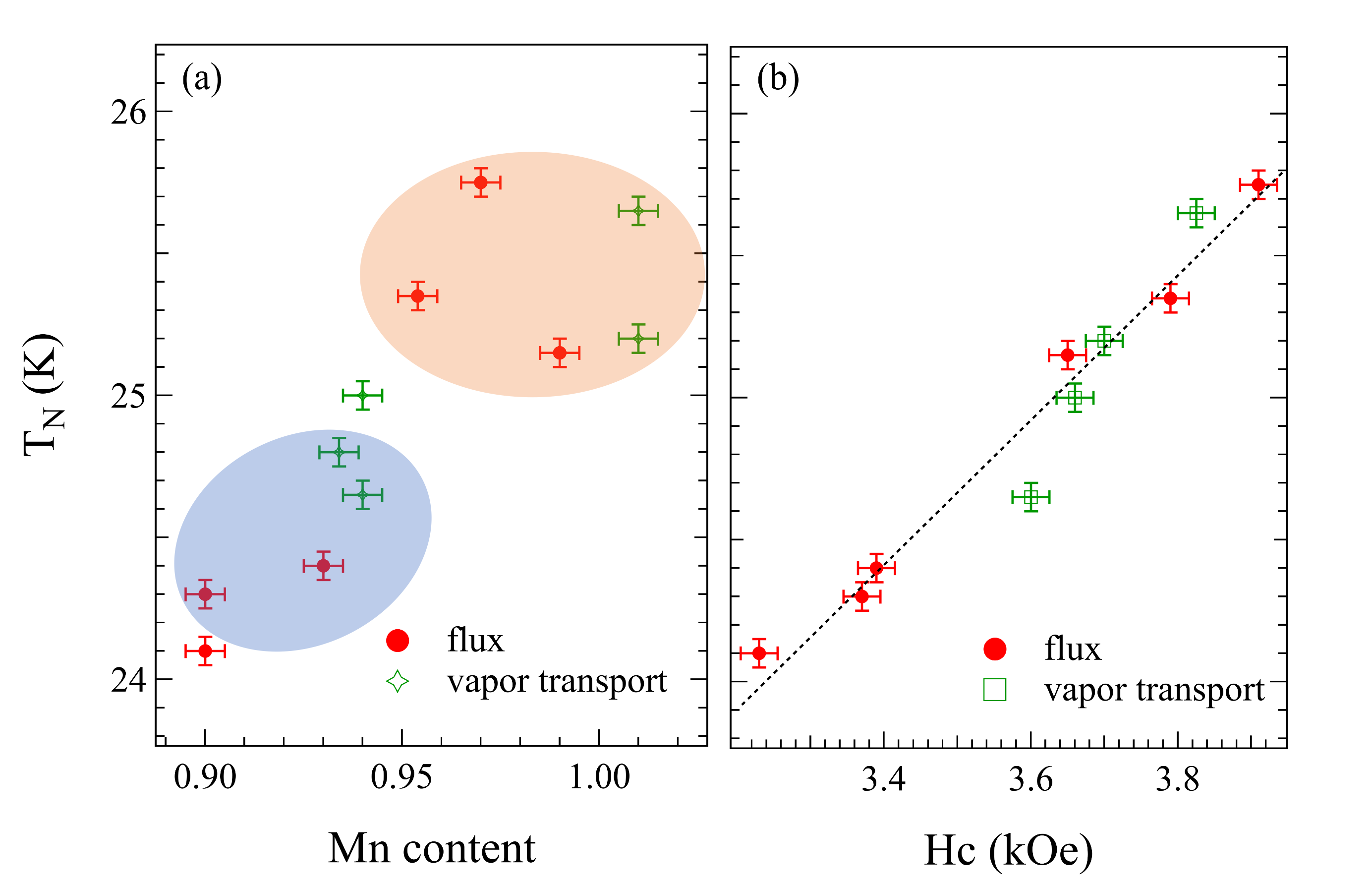}
\caption{(color online) (a) Dependence of the magnetic ordering temperature, T$_N$, on the Mn content. The upper circle shows $x_{Mn}$ range that are likely good for high temperature device performance. mK temperature is normally needed for devices using more Mn-deficient crystals (lower circle). (b) The relation between T$_N$ and the critical magnetic field, H$_c$, for spin flop transition. The Mn content was determined from EDS. T$_N$ was determined from temperature dependence of magnetization measured in an applied magnetic field of 1\,kOe.  H$_c$ was determined from field dependence of magnetization data collected at 2\,K.}
\label{TNMnHc-1}
\end{figure}

\subsection{Crystal screening before exfoliation}

\MBT~has been proved to be an ideal materials playground for investigating the intimate correlation between magnetism and electronic topology. There are ample experimental evidence that the device performance depends on the quality of starting bulk \MBT~single crystals. Considering the time consuming process of device fabrication and low temperature measurements, it is important to know what crystals could yield high quality flakes that would exhibit desired physical properties at the atomically thin limit. Ideally, one would be able to tell the crystal quality before exfoliation by routine bulk crystal characterization. Motivated by this, we have been trying to correlate the performance of thin flake devices from our collaborators\cite{ovchinnikov2021intertwined,cai2021electric} with the magnetic or transport properties of bulk crystals since these measurements can be easily performed in most advance laboratories. Figure\,\ref{TNMnHc-1}(a) shows the relation between T$_N$ and the Mn content, $x_{Mn}$, determined from elemental analysis. Here we consider the total Mn content only because it is too time consuming to determine the population and distribution of magnetic defects (Mn deficiency in the main Mn layer, and Mn$_{Bi}$ antisite defects) of every batch. Crystals with more Mn tend to have a higher T$_N$, but they don't exhibit a simple linear relation. The fact that T$_N$ varies even for crystals with the same $x_{Mn}$ signals the importance of magnetic defects. Our collabors showed that  quantized transport properties can be always observed above 2\,K  for flakes exfoliated from high quality crystals with $x_{Mn}>$0.94. In contrast, mK temperature is normally needed if more Mn deficient crystals are used. Figure\,\ref{TNMnHc-1}(b) shows the linear relation between T$_N$ and the critical field for spin flop transition, H$_{c1}$.  This linear relation indicates the importance of the interlayer exchange interaction on the magnetic properties and T$_N$ and H$_{c1}$  are equivalent in crystal screening. All \MBT~crystals that we synthesized order antiferromagentically  and have about 3\% Mn$_{Bi}$ which is not enough to change the sign of interlayer magnetic interactions. Therefore, T$_N$ and H$_{c1}$ can be employed to screen crystals before exfoliation and device fabrication.

\section{Summary}
In summary, we report the chemical vapor transport growth of \MBT, \MST, \MBST, and MnBi$_{4-x}$Sb$_x$Te$_7$ using I$_2$, MnI$_2$, MnCl$_2$, MoCl$_5$, and TeCl$_4$ as the transport agent. A small temperature gradient $<$20$^\circ$C is found to be critical for the crystallization of \MBT~and related compounds at the cold end of the growth ampoule. The vapor transported crystals grown with a small temperature gradient tend to have a stoichiometric Mn concentration. \MBT~crystals grown by both vapor transport and flux technique have similar amount of Mn$_{Bi}$. However, the chemical vapor transport technique leads to more antisite Mn ions in Sb-bearing compositions, as evidenced by the characterization of \MST, \MBST~and  MnBi$_{4-x}$Sb$_x$Te$_7$ crystals. Both vapor transport and flux growth techniques can be employed to grow high quality crystals of \MBT~and related compounds. They can be complimentary to each other to meet the different scientific needs. The difference between these two growth techniques provides us more options to fine tune the magnetic and topological properties via defect engineering.

This careful work is far from a thorough investigation of the vapor transport growth of \MBT~and related compounds. Instead, we hope this work would initiate more systematic studies on the vapor transport growths to further optimize the growth parameters for each member of the large family, a mission impossible for a single group. We would also hope that this work can motivate more studies on the fine tuning of the magnetic and topological properties of these fascinating compounds via materials synthesis. In addition to further optimizing the parameters of vapor transport growths, the following work deserves some special attention from crystal growth point of view.

(1) To identify the vapor transport conditions and develop the growth protocol for other $n$MnTe.$m$Bi$_2$Te$_3$ members.  The presence of regions with a composition of Mn$_2$Bi$_2$Te$_5$ in a \MBT~crystal is encouraging and indicates that vapor transport growth might be a valid route to sizable crystals of $n$MnTe.Bi$_2$Te$_3$ ($n>$1). The successful growth of MnBi$_{4-x}$Sb$_x$Te$_7$ crystals in this work occurs by serendipity instead of by design. Considering that vapor transported crystals tend to be less Mn deficient, a carefully optimized vapor transport growth might lead to high quality  MnTe.$m$Bi$_2$Te$_3$ ($m>$1) crystals for the study of their intrinsic properties and the theoretically predicted exotic properties. 

(2) To use the combination of two or more transport agents or to identify new transport agents.  Our work shows that iodides and chlorides are good transport agents for the growth of \MBT~and related compounds. We would expect that bromides or Br$_2$ are also valid transport agents and the appropriate growth temperature should also be in the range 500-590$^\circ$C which is not that far away from that used for flux growth. It is reasonable to expect that \MBT~crystals grown using bromides as transport agent might have similar amount of Mn$_{Bi}$. To further increase or reduce the amount of magnetic defects, using the combination of two or more transport agents might be a valid approach. It should be noted that a large number of halides, rather than only those studied in this work, can be effective transport agents. In addition to mixing halogen or halides, adding some C, S, Se, H$_2$ might be helpful. Identifying new transport agents or combinations might help expand the growth temperature window of \MBT~and facilitate the growth of  other $n$MnTe.$m$Bi$_2$Te$_3$ members.

(3) Possible low temperature growth in halide flux. The mechanism of growing \MBT~crystals good for QAHE out of MnCl$_2$ flux is still not clear. The vapor pressure of chlorides might play a role during the mass transport. Mixing MnCl$_2$ and other chloride salts can significantly lower the melting temperature and allow crystal growths at lower temperatures. Crystal growths of \MBT~and related compounds out of a mixture of halide salts can be performed in the conventional  vertical configuration or in the horizontal configuration\cite{yan2017flux}; the latter is ideal for a precise control of the precipitates in a system such as MnTe-Bi$_2$Te$_3$ with multi-phases in a narrow temperature range. Once the appropriate salt mixture is identified, the horizontal flux growth can be more effective in identifying the right growth temperature of each phase taking advantage of the temperature gradient along the ampoule.

\section{Acknowledgment}
The authors would thank Michael McGuire, Brian Sales, and Xiaodong Xu for helpful discussions. Work at ORNL was supported by the U.S. Department of Energy, Office of Science, Basic Energy Sciences, Materials Sciences and Engineering Division.

 This manuscript has been authored by UT-Battelle, LLC, under Contract No.
DE-AC0500OR22725 with the U.S. Department of Energy. The United States
Government retains and the publisher, by accepting the article for publication,
acknowledges that the United States Government retains a non-exclusive, paid-up,
irrevocable, world-wide license to publish or reproduce the published form of this
manuscript, or allow others to do so, for the United States Government purposes.
The Department of Energy will provide public access to these results of federally
sponsored research in accordance with the DOE Public Access Plan (http://energy.gov/
downloads/doe-public-access-plan).

\section{references}
\bibliographystyle{apsrev4-1}

%

\end{document}